# Cloud-based Big Data Analytics – A Survey of Current Research and Future Directions

Samiya Khan[1], Kashish Ara Shakil and Mansaf Alam
[1] samiyashaukat@yahoo.com
Department of Computer Science, Jamia Millia Islamia, New Delhi

## Abstract

The advent of the digital age has led to a rise in different types of data with every passing day. In fact, it is expected that half of the total data will be on the cloud by 2016. This data is complex and needs to be stored, processed and analyzed for information that can be used by organizations. Cloud computing provides an apt platform for big data analytics in view of the storage and computing requirements of the latter. This makes cloud-based analytics a viable research field. However, several issues need to be addressed and risks need to be mitigated before practical applications of this synergistic model can be popularly used. This paper explores the existing research, challenges, open issues and future research direction for this field of study.

**Keywords:** Cloud-based Big Data Analytics, Big Data, Big Data Analytics, Big Data Cloud Computing

## Introduction

With the advent of the digital age, the amount of data being generated, stored and shared has been on the rise. From data warehouses, webpages and blogs to audio/video streams, all of these are sources of massive amounts of data. The result of this proliferation is the generation of massive amounts of pervasive and complex data, which needs to be efficiently created, stored, shared and analyzed to extract useful information.

This data has huge potential, ever-increasing complexity, insecurity and risks, and irrelevance. The benefits and limitations of accessing this data are arguable in view of the fact that this analysis may involve access and analysis of medical records, social media interactions, financial data, government records and genetic sequences. The requirement of an efficient and effective analytics service, applications, programming tools and frameworks has given birth to the concept of Big Data Processing and Analytics.

Big data analytics has found application in several domains and fields. Some of these applications include medical research, solutions for the transportation and logistics sector, global security and prediction and management of issues concerning the socio-economic and environmental sector, to name a few. Apart from standard applications in business and commerce and society administration, scientific research is one of the most critical applications of big data in the real world [30]. According to O'Driscoll, Daugelaite and Sleator [32], one of the main future applications of big data analytics and cloud computing lies in life sciences. Some of the identified high-impact areas include systems biology, structure and protein function prediction, personalized medicine and metagenomics. Besides this, one of the most relevant applications of big data analytics is to improve the existing business models for efficiency and customer satisfaction.

Big data, by definition, is a term used to describe a variety of data - structured, semi-structured and unstructured, which makes it a complex data infrastructure [18]. The complexity of this infrastructure requires powerful management and technological solutions. One of the commonly used models for explaining big data is the multi-V model. Figure 1 illustrates the multi-V model.

Some of the Vs used to characterize big data include variety, volume, velocity, veracity and value [4]. The different types of data available on a dataset determine variety while the rate at which data is produced determines velocity. Predictably, the size of data is called volume. The two additional characteristics, veracity and value, indicate data reliability and worth with respect to big data exploitation, respectively.

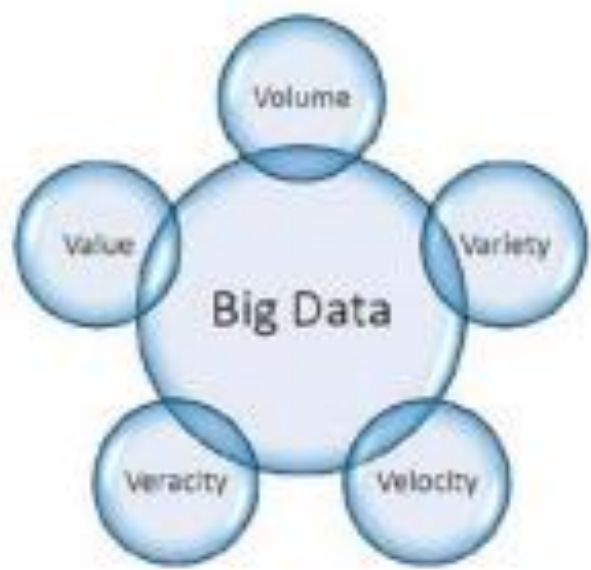

Fig. 1 – Big Data Characteristics [34]

In addition, Wu, Zhu, Wu and Ding [26] gave another characterization called the HACE theorem. According to this theorem, big data has two main characteristics. Firstly, it has a large volume of data that comes from different and heterogeneous sources, which is complex in nature. Secondly, the data is decentralized and distributed in nature.

Data is the central element of communication and collaboration in Internet and all the applications that are built on this platform. The immense popularity of data intensive applications like Facebook, LinkedIn, Twitter, Amazon, eBay and Google+ contributes to increasing requirement of storage and processing of data in the cloud environment. Schouten [21] uses Gartner's estimation to predict that by the year 2016, half of the data will be on the cloud.

Moreover, the data mining algorithms used for Big Data analytics possess high computing requirements. Therefore, they require high performance processors to do the job. The cloud provides a good platform for big data storage, processing and analysis, addressing two of the main requirements of big data analytics, high storage and high performance computing.

The cloud computing environment offers development, installation and implementation of software and data applications 'as a service'. Three multi-layered infrastructures namely, platform as a service (PaaS), software as a service (SaaS), and infrastructure as a service (IaaS), exist. Infrastructure-as-a-service is a model that provides computing and storage resources as a service. On the other hand, in case of PaaS and SaaS, the cloud services provide software platform or software itself as a service to its clients.

The cost of storage has considerably reduced with the advent of cloud-based solutions. In addition, the 'pay-as-you-go' model and the concept of commodity hardware allow effective and timely processing of large data, giving rise to the concept of 'big data as a service'. An example of one such platform is Google BigQuery, which provides real-time insights from big data in the cloud environment [12]. Shakil, Sethi, and Alam[37] demonstrates the application of cloud for management of Big Data in educational institutions which special focus on University-level data.

However, there have not been many practical applications of big data analytics that make use of the cloud. This has led to an increasing shift of research focus towards cloud-based big data analytics. An issue that is evident in this arrangement is information security and data privacy. As part of the cloud services, trust in data is also defined as a service. There shall be a considerable decrease in trust in view of the fact that the chances of security breaches and privacy violation will significantly rise upon implementation of big data strategies in the cloud. In addition, another important issue of ownership and control will also exist.

However, the potential of cloud-based big data analytics has compelled researchers to look into the existing issues to explore solutions. This paper discusses the different facets and aspects of data mining techniques/strategies adoption in the cloud environment for big data analytics. Moreover, it also looks into the existing research, identified challenges and future research directions in cloud-based big data analytics.

## BACKGROUND

Traditional data management tools and data processing or data mining techniques cannot be used for Big Data Analytics for the large volume and complexity of the datasets that it includes. Conventional business intelligence applications make use of methods, which are based on traditional analytics methods and techniques and make use of OLAP, BPM, Mining and database systems like RDBMS.

It was in the 1980s that artificial intelligence-based algorithms were developed for data mining. Wu, Kumar, Quinlan, Ghosh, Yang, Motoda, McLachlan, Ng, Liu, Yu, Zhou, Steinbach, Hand and Steinberg [25] mention the ten most influential data mining algorithms k-means, C4.5, Apriori, Expectation Maximization (EM), PageRank, SVM (support vector machine), AdaBoost, CART, Naïve Bayes and kNN (k-nearest neighbors). Most of these algorithms have been used commercially as well. Alam and Shakil [38] propose architecture for management of data through cloud techniques.

One of the most popular models used for data processing on cluster of computers is MapReduce. Jackson, Vijayakumar, Quadir and Bharathi [33] provide a survey on the programming models that support big data analytics. It identifies MapReduce/Hadoop as the most productive model for Big Data Analytics yet mentions that languages and extensions like HiveQL, Latin and Pig have overpowering benefits for this use.

Hadoop is simply an open-source implementation of the MapReduce framework, which was originally created as a distributed file system. According to Neaga and Hao [19], the evolution of Hadoop as a complete ecosystem or infrastructure that works alongside MapReduce components and includes a range of software systems like Hive and Pig languages, a coordination service called Zookeeper and a distributed table store called HBase.

For cloud-based big data analytics, several frameworks like Google MapReduce, Spark, Haloop, Twister, Hadoop Reduce and Hadoop++ are available. Figure 2 gives a pictorial representation of the use of cloud computing in big data analytics. These frameworks are used for storing and processing of data. In order to store this data, which may be of any structure, databases like HBase, BigTable and HadoopDB may be used. When it comes to data processing, the Pig and Hive technologies come into the picture.

Some of the recent research breakthroughs and milestones in cloud-based big data analytics are discussed here. Lee [16] elaborates on the advantages and limitations of MapReduce in parallel data analytics. A Hadoop-based data analytics system, created by Starfish [13], improves the performance of the clusters throughout the cycle of data analytics. Moreover, the users are not required to understand the configuration details.

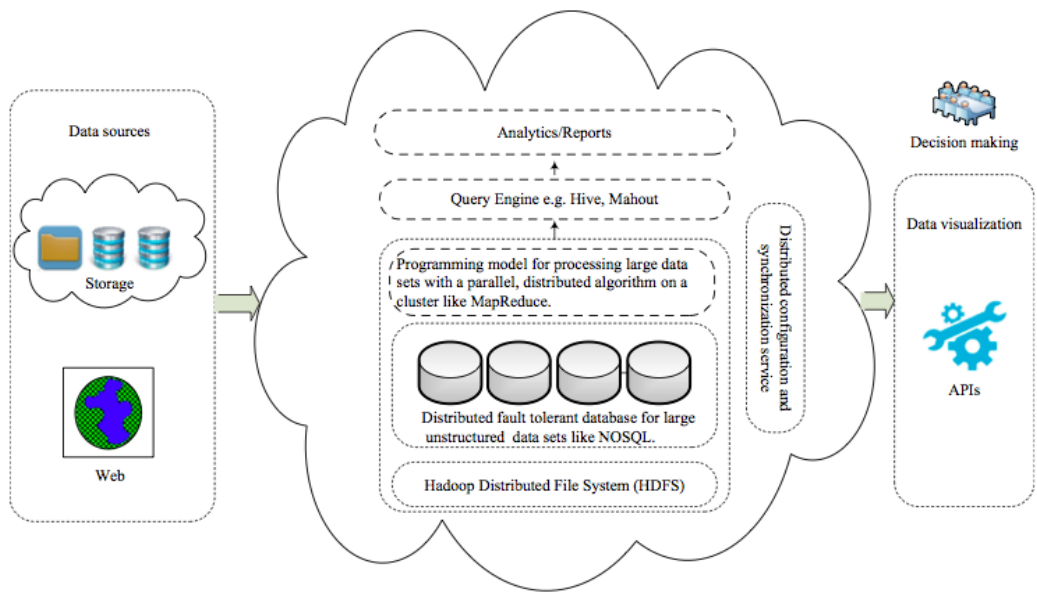

Fig. 2 – Use of Cloud Computing in Big Data [27])

In recent times, lack of interactivity has been identified as a major issue and several efforts have been made in this area. Borthakur, Gray, Sarma, Muthukkaruppan, Spiegelberg, Kuang, Ranganathan, Molkov, Menon, Rash, Schmidt and Aiyer [5] optimize the HBase and HDFS implementation for better responsiveness. Strambei [23] evaluates the viability of OLAP Web Services for cloud-based architectures, with the specific objective to allow open and wide access to web analytical technologies.

Research efforts have been made to create a big data management framework for the cloud. Khan, Naqvi, Alam and Rizvi [35] propose a data model and provides a schema for big data in the cloud and attempts to ease the process of querying data for the user. Moreover, an important subject of research has been performance and speed of operation. Ortiz, Oneto and Anguita [28] explore the use of a proposed integrated Hadoop and MPI/OpenMP system and how the same can improve speed and performance.

In view of the fact that data needs to be transferred between data centers that are usually located distances apart, power consumption becomes a crucial parameter when it comes to analyzing efficiency of the system. A network-based routing algorithm called GreeDi can be used for finding the most energy efficient path to the cloud data center during big data processing and storage [29].

There are several practical simulation-enabled analytics systems. One such system is given by Li, Calheiros, Lu, Wang, Palit, Zheng and Buyya [17], which is a Direct Acrylic Graph (DAG) form analytical application used for modeling and predicting the outbreak of Dengue in Singapore.

Online risk analytics and the need for an infrastructure that can provide users the programming resources and infrastructure for carrying out the same have also appeared in the form of Aneka [6] and CloudComet [15]. Chen [7] investigates the concept of CAAAS or Continuous Analytics As A Service, which is used for predicting the behavior of a service or a user.

The last topic under Big Data Analysis that has caught the attention of the research community is Real-time Big Data Analysis. Many commercial cloud service providers are providing solutions for real-time analysis. AWS based-solutions for real-time stream processing is AWS Kinesis [2]. Many frameworks and software systems have also been introduced for this purpose, some of which are Apache S4 [3], IBM InfoSphere Streams [14] and Storm [22].

## CHALLENGES AND ISSUES

In order to move beyond the existing techniques and strategies used for machine learning and data analytics, some challenges need to be overcome. NESSI [20] identifies the following requirements as critical.

1. In order to select an adequate method or design, a solid scientific foundation needs to be developed.
2. New efficient and scalable algorithms need to be developed.
3. For proper implementation of devised solutions, appropriate development skills and technological platforms must be identified and developed.
4. Lastly, the business value of the solutions must be explored just as much as the data structure and its usability.

In view of cloud-based big data analytics, additional challenges like adoption and implementation of effective big data solutions using cloud architecture and mitigating the security and privacy risks also exist. One of the biggest concerns while using big data analytics and cloud computing in an integrated model is security. This is perhaps the reason why this aspect of cloud-based big data analytics and its practical usage and implementation has attracted immense attention.

Liu, C., Yang, C., Zhang, X. and Chen, J. [31] provides a summary, analysis and comparison of authenticator-based data integrity verification techniques on cloud and Internet-of-things data. This paper suggests that any future developments in this area needs to look at three main aspects namely, efficiency, security and scalability/elasticity.

A G-Hadoop based security framework is proposed by Zhao, Wang, Tao, Chen, Sun, Ranjan, Kołodziej, Streit and Georgakopoulos [36], which makes use of solutions like SSL and public key cryptography for ensuring security of big data resident on distributed cloud data centers. In addition to several security mechanisms, this framework also aims to simplify the processes of submitting job and authenticating users.

Talia [24] suggests further research and development in the following areas:

1. Programming abstracts or scalable high-level models and tools.
2. Solutions for data and computing interoperability issues.
3. Integration of different big data analytics frameworks
4. Techniques for mining provenance data

## FUTURE RESEARCH DIRECTIONS

Several open source data mining techniques, resources and tools exist. Some of these include R, Gate, Rapid-Miner and Weka, in addition to many others. Cloud-based big data analytics solutions must provide a provision for the availability of these affordable data analytics on the cloud so that cost-effective and efficient services can be provided. The fundamental reason why cloud-based analytics are such a big thing is their easy accessibility, cost-effectiveness and ease of setting up and testing. In view of this, some of the main research directions identified by Neaga and Hao [19] include:

1. Evolution of analytics and information management with respect to cloud-based analytics.
2. Adaptation and evolution of techniques and strategies to improve efficiency and mitigate risks.
3. Formulate strategies and techniques to deal with the privacy and security concerns.

4. Analysis and adaptation of legal and ethical practices to suit the changing viewpoint, impact and effects of technological advances in this regard.

With this said, the research directions are not limited to the above-mentioned points. The main goal is to transform the cloud from being a data management and infrastructure platform to a scalable data analytics platform.

## CONCLUSION

This is an age of big data and the emergence of this field of study has attracted the attention of many practitioners and researchers. Considering the rate at which data is being created in the digital world, big data analytics and analysis have become all the more relevant. Moreover, most of this data is already on the cloud. Therefore, shifting big data analytics to the cloud framework is a viable option.

Moreover, the cloud infrastructure suffices the storage and computing requirements of data analytics algorithms. On the other hand, open issues like security, privacy and the lack of ownership and control exist. Research studies in the area of cloud-based big data analytics aim to create an effective and efficient system that addresses the identified risks and concerns.